\documentclass[apj]{emulateapj} \usepackage{times} 

\usepackage{graphicx}
\usepackage{amssymb}
\usepackage{amsmath}
\usepackage{natbib}
\usepackage{times} 
\usepackage{epstopdf}

\usepackage{calc} 

\def\be{\begin{equation}}
\def\ee{\end{equation}}
\def\lsim{\raise0.3ex\hbox{$\;<$\kern-0.75em\raise-1.1ex\hbox{$\sim\;$}}}
\def\gsim{\raise0.3ex\hbox{$\;>$\kern-0.75em\raise-1.1ex\hbox{$\sim\;$}}}

\newcommand{\phant}{\phantom{0}}

\begin{document}

\title{New calculation of antiproton production by cosmic ray protons
and nuclei}

\author{Michael Kachelrie\ss\altaffilmark{1}, Igor V.~Moskalenko\altaffilmark{2} and Sergey S.~Ostapchenko\altaffilmark{2,3}}

\affil{$^1$Institutt for fysikk, NTNU, 7491 Trondheim, Norway}
\affil{$^2$Hansen Experimental Physics Laboratory \& Kavli Institute for Particle Astrophysics and Cosmology, Stanford University, Stanford, CA 94305, U.S.A}
\affil{$^3$Skobeltsyn Institute of Nuclear Physics, Moscow State University, 119991 Moscow, Russia}

%%%%%%%%%%%%%%%%%%%%%%%%%%%%%%%%%%%%%%%%%%%%%%%%%%%%%%%%%%%%%%%%%%%%%%%%%%%%
\begin{abstract}
A dramatic increase in the accuracy and statistics of space-borne cosmic ray (CR) measurements has yielded several breakthroughs over the last several years.
The most puzzling is the rise in the positron fraction above $\sim$10\,GeV over the predictions of the propagation models assuming pure secondary production.
The accuracy of the antiproton production cross section is critical for 
astrophysical applications and searches for new physics since
antiprotons in CRs seem to hold the keys to many puzzles including the origin of those excess positrons. 
However, model calculations of antiproton production in CR interactions with interstellar gas
 are often employing parameterizations 
that are out of date or 
are using outdated physical concepts. That may lead to an incorrect interpretation of antiproton data
which   could have broad consequences for other areas of astrophysics. In this work, we calculate antiproton production in $pp$-, 
$pA$-, and $AA$-interactions using EPOS-LHC and QGSJET-II-04, two of the most 
advanced Monte Carlo (MC) generators tuned to numerous accelerator data
including those from the Large Hadron Collider (LHC). We show that the antiproton yields 
obtained with these MC generators differ by up to an order of magnitude from yields of 
parameterizations commonly used in astrophysics.
\end{abstract}


%%%%%%%%%%%%%%%%%%%%%%%%%%%%%%%%%%%%%%%%%%%%%%%%%%%%%%%%%%%%%%%%%%%%%%%%%%%%
\section{{Introduction}\label{intro.sec}}

%Antiprotons in CRs attract a lot of attention since their discovery. 
Antiprotons in cosmic rays (CR) are produced in CR interactions with interstellar gas and are, therefore,
called secondary. The same interactions produce charged and neutral mesons that decay to secondary electrons and positrons and $\gamma$-rays.
However, in contrast to CR positrons that can be produced copiously in pulsars, there is no known
astrophysical source of primary antiprotons.
Since the antiproton ``background'' is fairly low this provides an opportunity to search for
a new phenomenon or an exotic signal if the CR antiproton flux is measured accurately.
An even smaller background is expected for
 antideuterons, which provide, therefore,
another target in searches for new physics \citep{1997PhLB..409..313C,2000PhRvD..62d3003D}. 

The first detections of antiprotons in CRs were reported 35 years ago by \citet{1979ICRC....1..330B} and \citet{1979PhRvL..43.1196G}.
Both experiments reported
 a flux that was a factor of $\sim$3 above expectations \citep{1974PhRvD..10.1731G} and caused a stir
at that time even though the reported excess was only at the $\sim$$3\sigma$ level. 
Now it is clear that these measurements and a followup measurement by \citet{1981ApJ...248.1179B} were likely plagued by
a background of negatively charged particles. Meanwhile, 
these measurements stimulated an intensive discussion about the possiblity
to observe primary antiprotons produced in exotic processes, such as the annihilation of primordial black holes or dark matter in the Galactic halo
\citep{1980Natur.285..386S,1981Natur.293..120K,1984PhRvL..53..624S,1985otar.conf....1C,1988ApJ...325...16R,1988PhLB..214..403E,1989ApJ...336L..51S,1994PhRvD..49.2316J,1996PhLB..384..161C,1996PhR...267..195J,1996PhRvL..76.3474M}.

Subsequent flights by the MASS91 \citep{1996ApJ...467L..33H}, CAPRICE98 \citep{2001ApJ...561..787B}, and most notably the BESS and BESS-Polar (1993-2008) 
\citep{2000PhRvL..84.1078O,2012PhRvL.108e1102A} experiments made a set of accurate measurements
that motivated also new attempts to make more accurate theoretical evaluations of the flux of secondary antiprotons produced in CR interactions with interstellar gas. 
As a result, it has become clear that the measured antiproton flux and $\bar p/p$ ratio are generally consistent with the secondary antiproton production 
\citep{1992ApJ...394..174G,1998A&A...338L..75M,1998ApJ...499..250S,1998PhRvD..58l3503B,1999PhRvL..83..674B,1999ApJ...526..215B,2001ApJ...563..172D,2002ApJ...565..280M}, but do not rule out the possibility of a weak exotic signal. 

More realistic calculations
involving the fully numerical CR propagation model GALPROP \citep{1998ApJ...509..212S,2007ARNPS..57..285S} have shown that the predicted flux depends on the assumed propagation model 
\citep{2002ApJ...565..280M},
even though the propagation
parameters were derived from fits to the B/C and $^{10}$Be/$^9$Be ratios and the CR fluxes were tuned to  local measurements \citep{2007ARNPS..57..285S}.
In particular, a standard stochastic reacceleration model was found to underproduce antiprotons by $\sim$40\% at 3\,GeV, while a semi-empirical plain diffusion model 
overproduces them by $\sim$20\%. A diffusion-convection model was found to be consistent with BESS 95-97 data. 
%For all these models propagation
%parameters were derived from the fits to the B/C and $^{10}$Be/$^9$Be ratios and CR fluxes were tuned to the local measurements.
The origin of the remarkable discrepancy with the predictions of the physically motivated reacceleration model remained unclear, but
it was natural to blame the model deficiency rather than to declare an exotic signal.
% since all secondaries
%in CRs, including boron and antiprotons, should be reproduced automatically once the model parameters are fixed from the fit.
One of the possibilities to reconcile the model predictions with the data is to assume that 
some fraction of carbon and other primary nuclei at low energies is local \citep{2003ApJ...586.1050M}. In this case the
appropriate decrease in the derived diffusion coefficient increases the production of antiprotons, enough to reproduce the data.
Another solution is to account for the back reaction of CR protons onto the interstellar turbulence \citep{2006ApJ...642..902P}. 
The stochastic acceleration of CRs by MHD waves is accompanied by the so-called damping of the waves, since the wave energy is dissipated.
This also leads to the effective decrease in the diffusion coefficient at a few GV that increases the antiproton production.
 
Recent discoveries in astrophysics of CRs, however, have changed the landscape dramatically.
The most important was a clear confirmation of the rise\footnote{Relative to expectations as if all positrons in CRs are secondary \citep{1982ApJ...254..391P,1998ApJ...493..694M}.} 
in the positron fraction by PAMELA \citep{2009Natur.458..607A,2013PhRvL.111h1102A} that was earlier 
noticed by the HEAT collaboration \citep{2004PhRvL..93x1102B}. This was subsequently confirmed and extended to higher energies by \emph{Fermi}-LAT \citep{2012PhRvL.108a1103A}, and measured with even greater precision and extended to even higher energies by AMS-02 \citep{2013PhRvL.110n1102A}. 
Most recently, the measurement of the positron fraction has been extended up 
to 500\,GeV by the AMS-02 collaboration hinting for some flattening of the
fraction above 200\,GeV \citep{2014PhRvL.113121101A}, 
but the statistics is still low in this energy range.
New accurate antiproton measurements were done by 
PAMELA \citep{2009PhRvL.102e1101A,2010PhRvL.105l1101A,2013JETPL..96..621A} 
covering the range from 70\,MeV -- 200\,GeV, and more is expected from AMS-02.
Above a few GeV, the data are consistent with secondary production, 
in strong contrast with positron results.
Another important milestone is an accurate measurement of the B/C ratio up 
to 100\,GeV/nucleon by PAMELA \citep{2014ApJ...791...93A} that is consistent
with preliminary 
AMS-02\footnote{http://www.ams02.org/2013/07/new-results-from-ams-presented-at-icrc-2013/ \label{foot1}} 
measurements reaching $\sim$400\,GeV/nucleon. 
Both measurements indicate   the index of the diffusion coefficient
$\alpha\approx 0.4$ or even smaller \citep{2014ApJ...791...93A} that
supports a Kolmogorov-type power spectrum of interstellar 
turbulence \citep{1941DoSSR..30..301K}, 
thus favoring the stochastic reacceleration model for interstellar propagation. 
Besides providing more accurate data over a wider energy range these new measurements give an important inside
into CR acceleration and propagation processes. Also relevant, but controversial are
the new measurements of CR proton and He spectra, the most abundant CR species. 
Combined measurements by PAMELA \citep{2011Sci...332...69A}, ATIC \citep{2009BRASP..73..564P}, and CREAM \citep{2010ApJ...714L..89A}
hint on a break in both spectra at about the same rigidity $\sim$230 GV. The flattening of the CR proton spectrum at high energies was 
also confirmed by the {\it Fermi}-LAT through observations of $\gamma$-ray emission of the Earth's limb \citep{2014PhRvL.112o1103A}. However,
preliminary results by AMS-02$^{\ref{foot1}}$ do not show any feature in the CR proton and He spectra up to $\sim$2\,TV.

Antiproton data and their correct interpretation hold the key to the resolution of many astrophysical puzzles. 
If the rise in the positron fraction is due to WIMP annihilations, antiproton data provide important constraints on
WIMP models \citep{2009PhRvL.102g1301D,2009NuPhB.813....1C}, for a review, see \citet{2011ARA&A..49..155P}. 
If the rise is due to conventional astrophysics, antiproton and B/C measurements extended to 
higher energies may be able to discriminate between the pulsar 
\citep{1981IAUS...94..175A,1987ICRC....2...92H,1989ApJ...342..807B,2009JCAP...01..025H,2014JCAP...04..006D}
and SNR hypotheses \citep{2003A&A...410..189B,2009PhRvL.103e1104B}. 
The latter proposes a secondary component with a hard energy spectrum that is 
produced in a SNR shock by accelerated protons. 
It also predicts a rise in all secondaries, 
such as the $\bar p/p$ and B/C ratios, at high energies
\citep{2003A&A...410..189B,2009PhRvL.103h1103B,2014PhRvD..89d3013C,2014PhRvD..90f1301M}. 
However, other authors \citep{2011ApJ...733..119K,2013PhRvD..87d7301K} pointed out that the results of their
time-dependent Monte Carlo simulations predict a 
flattening rather than a distinct rise.
%However, such a rise is absent in the results of the Monte Carlo
%simulations performed by \citet{2011ApJ...733..119K} and 
%\citet{2013PhRvD..87d7301K} and
%they argued that the ``re-acceleration mechanism'' proposed in \citet{2009PhRvL.103e1104B} works only choosing physically inconsistent parameters.}

The accuracy of the antiproton production cross section is critical for 
astrophysical applications and searches for new physics. 
 This is especially true in view of many expectations connected with the upcoming data releases
by the AMS-02 experiment operating at the International Space Station, 
and by soon-to-be-launched ISS-CREAM\footnote{http://cosmicray.umd.edu/iss-cream/}, CALET\footnote{http://calet.phys.lsu.edu} and GAPS\footnote{http://gamma0.astro.ucla.edu/gaps}experiments.
This holds even more for new opportunities that would open up with antideuteron detection in CRs. 
In turn, the calculation of antideuteron production relies on the inclusive antiproton 
production cross sections and the detailed knowledge of 
two-particle correlations \citep{2010PhLB..683..248K,2012PhRvD..86j3536D}.

In this work, we analyze antiproton production in $pp$-, \mbox{$pA$-,} 
and $AA$-interactions using EPOS-LHC and QGSJET-II-04, two of the most 
advanced Monte Carlo (MC) generators tuned to numerous accelerator data 
including those from the Large Hadron Collider (LHC). 
The antiproton yields 
obtained with these MC generators differ 
by a factor of few from yields of parameterizations commonly 
used in astrophysics. 
The article is structured as follows: In Section~\ref{sec:models}, we compare
the pros and cons of parameterizations and MC generators developed
specifically for CR interactions at low- and high-energies.
In Section~\ref{sec:LE}, we present a tune of the fragmentation procedure used
in the QGSJET-II-04 model which leads to an improved description of particle
production at low energies
in the presence of high-energy thresholds, as in the case of  antiprotons.
Then we compare in Section~\ref{sec:zfactors} the results of the modified QGSJET-II-04 
and the EPOS-LHC models to experimental data and to the predictions
of the parameterizations of \citet{1983JPhG....9.1289T} and \citet{2003PhRvD..68i4017D}. 
Finally, we discuss the nuclear enhancement of the $\bar p$ yield
by nuclear species in Section~\ref{sec:enh} before we conclude.

%%%%%%%%%%%%%%%%%%%%%%%%%%%%%%%%%%%%%%%%%%%%%%%%%%%%%%%%%%%%%%%%%%%%%%%%%%%%
\section{$\bar{p}$ production: models and parameterizations} \label{sec:models}

Calculations of secondary CR fluxes, both for astrophysical applications
and searches for new physics, are typically based on empirical 
parameterizations of accelerator data for the production spectra of 
secondary particles in proton-proton interactions. Despite the evident 
convenience of using such parameterizations, this practice
has a number of caveats. First of all, one has to rely on empirical
scaling laws when these parameterizations are extrapolated outside of the
kinematic range of the data they are based on. In particular, 
the high energy extrapolations of existing parameterizations prove to be 
%highly 
unreliable, as we will demonstrate in the following. 
This failure partly reflects the scarcity of relevant experimental
data and thus the poor theoretical understanding of the high energy behavior 
of hadronic collisions at the time when the scaling laws used as input 
were developed. Secondly, it is problematic to account for the contributions
of nuclear species in CRs and/or in the interstellar medium
(ISM) using such parameterizations: While particle production in proton-nucleus 
collisions has been studied by a number of fixed target accelerator experiments,
yielding the spectra of secondaries in the forward hemisphere in the center-of-mass system (c.m.s.), there is practically no experimental
information on the forward particle spectra in nucleus-proton and
nucleus-nucleus collisions. One usually attempts to solve the problem
by introducing empirical ``nuclear enhancement'' factors, which
vary, however, substantially %(by as much as 50\%) 
between different publications.
Moreover, as it has been demonstrated in \citet{2014ApJ...789..136K}, properly
calculated nuclear enhancement factors depend strongly both on the
production spectra of the respective particles and on the spectral
slopes of CR species.

These difficulties fully apply to the case of antiproton production by CRs, 
which motivated us to reconsider the problem employing MC generators of 
hadronic interactions. Comparing the results of a number of MC models 
used in low energy nuclear physics to available 
%experimental 
accelerator data, we 
observed strong (up to an order of magnitude) deviations from
the  measured  $\bar{p}$ spectra. Hence, we decided to turn to MC generators of
high energy interactions, notably, to those used in the high energy 
CR field \citep{1997NuPhS..52...17K,2009PhRvD..80i4003A,2006PhRvD..74a4026O,2006PhRvC..74d4902W}.
Our choice is motivated by the fact that such models have been calibrated
on numerous accelerator data over a wide energy range and survived
many consistency tests in high energy CR studies. %\cite{kaskade99,01,05,09}.
Moreover, most of those models are in a good agreement
with various data from LHC Run I 
\citep{2011APh....35...98D} and some of them have been recently updated
\citep{2011PhRvD..83a4018O,2013EPJWC..5202001O,2013arXiv1306.0121P1}, 
including a re-tuning of model parameters, based on LHC data.

However, the application of such MC generators to low energy hadronic
interactions which dominate the production of secondary $\bar{p}$ 
in the physically important range of kinetic energies 
$E_{\bar p}^{{\rm kin}} \lsim 10$\,GeV is rather unwarranted. 
Such low energies imply an extrapolation
of the underlying theoretical approach outside of its range of applicability:
Several reaction mechanisms like contributions of Reggeon
exchanges, intranuclear cascading, etc., which are important at low energies
% (for a review see \cite{eng06})
are irrelevant for high energy interactions and are therefore typically 
neglected. Comparing the model predictions with
experimentally measured antiproton spectra for proton-proton and proton-nucleus
collisions for incident momenta $p_{{\rm lab}}\lesssim100$\,GeV/c,
we observed generally a strong disagreement with data%
\footnote{Similar discrepancies have been reported in \citet{2013PhRvD..88b3014I}
for the case of the DPMJET-III model \citep{2001amcr.conf.1033R} which is employed
in the popular FLUKA code \citep{2007AIPC..896...31B}.}. 
In the particular case of the QGSJET-II-04 model \citep{2011PhRvD..83a4018O,2013EPJWC..5202001O},
this disagreement was especially large close to the kinematical
threshold for $\bar{p}$ production. Only the EPOS-LHC model \citep{2013arXiv1306.0121P1}
demonstrated a generally reasonable behavior in the low energy limit,
as will be demonstrated in the following.

%%%%%%%%%%%%%%%%%%%%%%%%%%%%%%%%%%%%%%%%%%%%%%%%%%%%%%%%%%%%%%%%%%%%%%%%%%%
\section{Low energy extension of the QGSJET-II model} \label{sec:LE}

%As discussed in the previous Section, t
The results of EPOS-LHC and 
QGSJET-II-04 for $\bar{p}$ production are rather similar in the high 
energy range where both MC generators are tuned to LHC data,
but deviate strongly in the low energy limit. 
At first sight, this seems surprising as their treatment of hadronic 
interactions at relatively low collision energies is very similar.
The common underlying physics 
%picture is 
includes
%the one of 
multiple scattering processes, 
which are described within the Reggeon Field Theory framework \citep{1968JETP...26..414G} 
by multiple exchanges of Pomerons, i.e.\ composite states with vacuum quantum 
numbers. Particle production is treated as the formation and break-up of 
strings of color field, which is performed using string fragmentation. 
However, the procedures used in the two models to fragment color strings differ,
and these differences are responsible for the discussed
variations at low energies. Since QGSJET-II-04 reproduces well
pion and photon production data down to rather low energies, 
$p_{{\rm lab}}$$\sim$$10$\,GeV/c \citep{2012PhRvD..86d3004K}, it is natural to suspect that 
the discrepancies in $\bar p$ production are caused by threshold effects 
related to the relatively high antiproton mass.

These observations motivated us to improve the low energy behavior of
QGSJET-II, modifying its string fragmentation procedure. It is worth 
stressing that we did not
aim  at the development of
a full-scale model for low energy hadronic collisions.
Instead, we intended a more reliable extrapolation of the current model
towards low energies---in order to reach an acceptable agreement with 
experimental data in the energy range relevant for calculations of the
CR antiproton flux. In particular, the modifications introduce
no additional adjustable parameters. They only modify slightly
the string fragmentation algorithm, in such a way that the changes
have no significant influence on the results in the high energy range. 

To describe these modifications in some more detail, let us briefly 
discuss the hadronization procedure used in QGSJET-II. The standard physics 
picture for the string break-up is the neutralization of the color field 
via the creation
of quark-antiquark and diquark-antidiquark pairs from the vacuum, followed by
their conversion into final-state hadrons~\citep{1983PhR....97...31A}. In
QGSJET-II, this process is modelled using an iterative procedure~\citep{1993PAN....56..346K},
whose parameters are determined by the intercepts of the corresponding
Regge trajectories~\citep{kai87}. After creating a new quark-antiquark
(diquark-antidiquark) pair, a hadron is formed and the reminder of
the string proceeds to the next step: Depending on the string mass,
either the same procedure is repeated or a two-particle decay is modeled.
In the new treatment, we introduced an additional weight into the
sampling process of kinematic variables for each subsequent string break-up,
which is proportional to the two-particle decay phase volume evaluated 
for the mass squared of the reminder of the string.
While being of minor importance for high mass strings,
stretched over long rapidity intervals, which are typically produced in
 high energy collisions, this modification enhances the production of 
 light hadrons (mostly pions) in the fragmentation of strings of small
 masses, at the expense of the heavier hadrons.
 Additionally, the parameters of the hadronization procedure, notably
 the string mass threshold for proceeding to a two-particle decay and
 the relative probabilities for creating quark-antiquark and 
 diquark-antidiquark pairs from the vacuum, have been readjusted in order
 to stay in agreement with high energy data.

\begin{figure*}[p]
\center{
\includegraphics[width=0.9\textwidth]{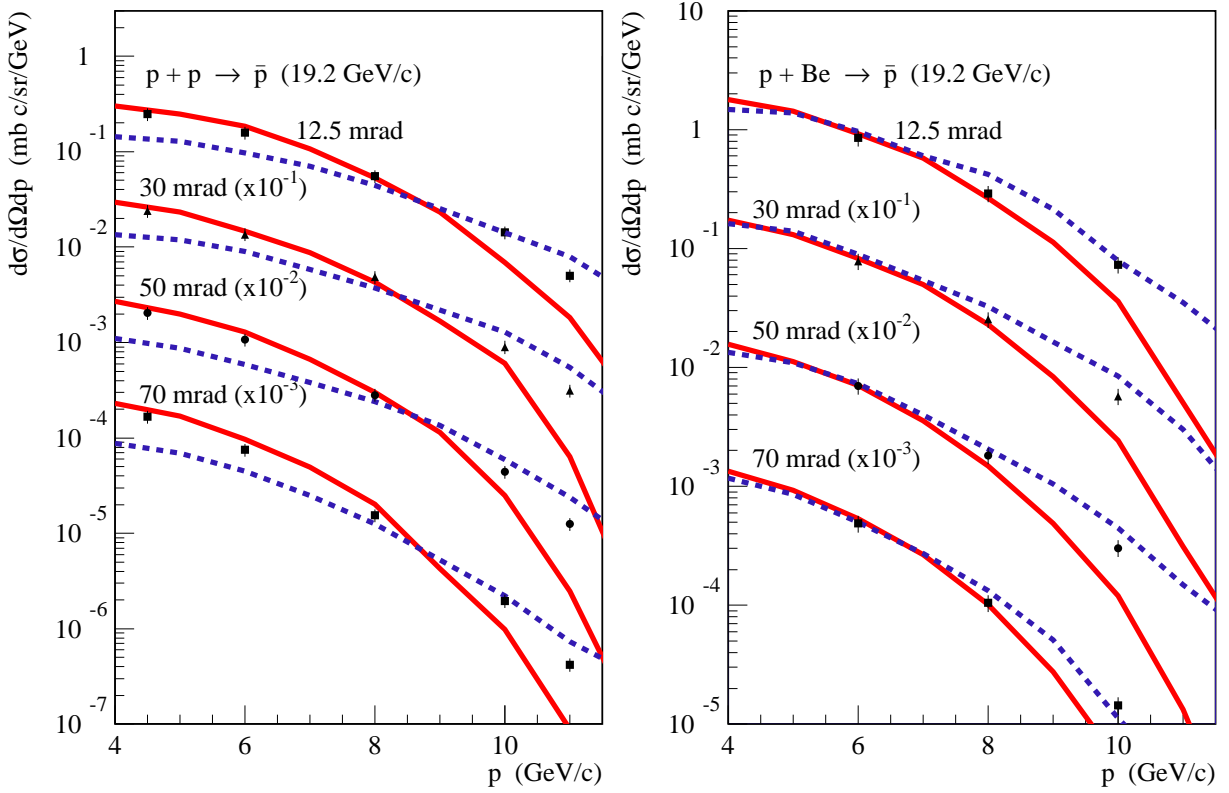}
}
\vskip5mm
\caption{Momentum spectra of $\bar p$'s in the laboratory frame in $pp$-collisions (left) and $p\,{\rm Be}$-collisions (right) 
at $p_{\rm lab}=19.2$,GeV/c, for different angles $\theta$ in the laboratory frame
(as indicated in the plots), calculated using QGSJET-IIm (solid, red) and
EPOS-LHC (dashed, blue). The experimental data -- \citet{allaby70}
 and \citet{1975NuPhB..86..403A}.\label{fig:ap19}}
\end{figure*}

\begin{figure*}[p]
\center{
\includegraphics[width=0.8\textwidth]{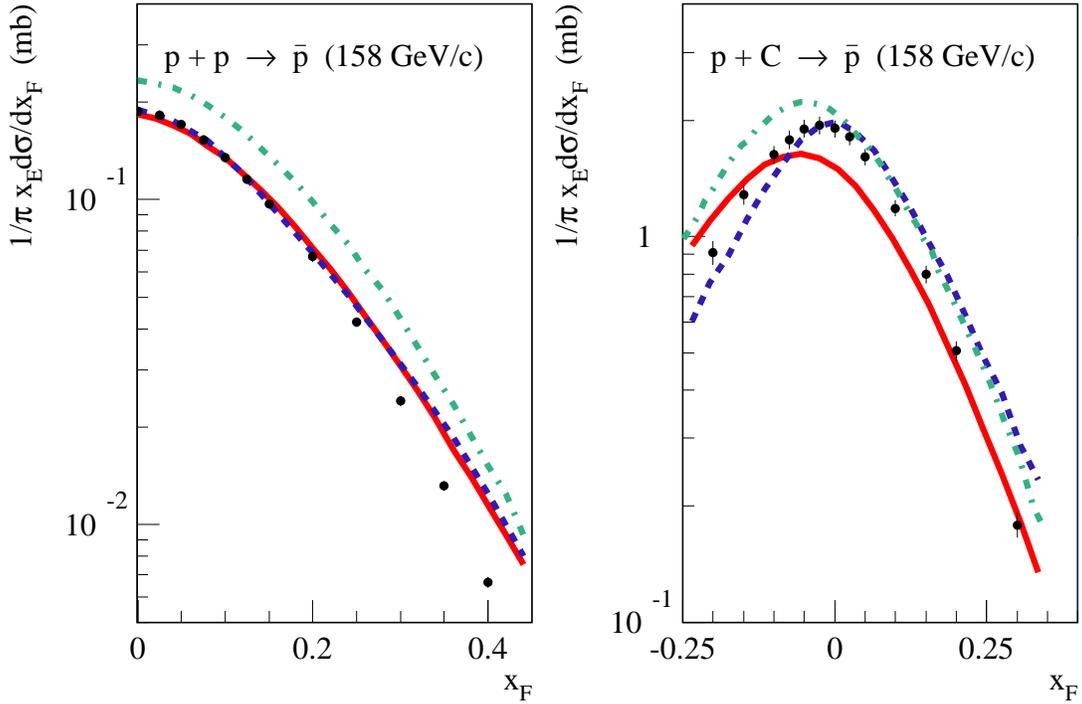}
}
\caption{Feynman $x$-spectra of antiprotons, $1/\pi\,x_{\rm E}\,d\sigma/dx_{\rm F}$ (see text for definition),
%($x_{\rm E}=2E^*/\sqrt{s}$, $x_{\rm F}=2p_z^*/\sqrt{s}$, with $E^*$ and
% $p_z^*$ being
%the c.m.s.\ energy and the $z$-component of the momentum) 
 in $pp$ (left) and
$p\,{\rm C}$ (right)
collisions at $p_{\rm lab}=158$\,GeV/c, calculated using
QGSJET-IIm (solid, red),
EPOS-LHC (dashed, blue), and QGSJET-II-04  (dot-dashed, green),
in comparison with NA49
data \citep{2010EPJC...65....9A,2013EPJC...73.2364B}.\label{fig:ap158}}
\end{figure*}

\begin{figure}[p]
\center{
\includegraphics[width=0.45\textwidth,height=6cm]{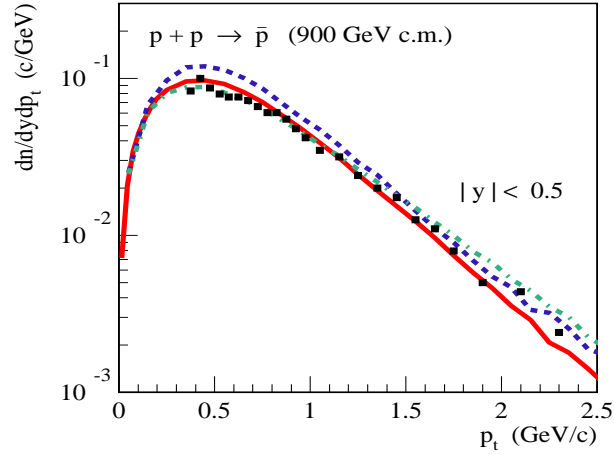}
}
\caption{Transverse momentum spectra of $\bar p$'s at central rapidity in c.m.s.\
($|y|<0.5$) in $pp$-collisions at $\sqrt s=900$\,GeV, calculated using 
QGSJET-IIm (solid, red),
EPOS-LHC (dashed, blue), and QGSJET-II-04  (dot-dashed, green),
in comparison with ALICE
data \citep{2011EPJC...71.1655A}.\label{fig:ap900}}
\end{figure}

In Figs.~\ref{fig:ap19} and \ref{fig:ap158}, 
we compare the results of the modified QGSJET-II model (hereafter referred to as
QGSJET-IIm) and the EPOS-LHC model
with selected benchmark accelerator data\footnote
{A comparison with additional experimental data sets is presented
in Appendix~\ref{Ap-A}.}. The data shown in Fig.~\ref{fig:ap19} are the
momentum spectra of $\bar p$ in the laboratory frame in $pp$-collisions 
and $p$\,Be-collisions
at $p_{\rm lab}=19.2$\,GeV/c \citep{allaby70,1975NuPhB..86..403A} for different
 angles $\theta$ in the laboratory frame.
The data shown in Fig.~\ref{fig:ap158} are the Feynman $x$-spectra of antiprotons,
 $1/\pi\,x_{\rm E}\,d\sigma/dx_{\rm F}$, where
$x_{\rm E}=2E^*/\sqrt{s}$, $x_{\rm F}=2p_z^*/\sqrt{s}$, with $E^*$ and
 $p_z^*$ being
the c.m.s.\ energy and the $z$-component of the momentum,
in $pp$- and $p$\,C-collisions at $p_{\rm lab}=158$\,GeV/c \citep{2010EPJC...65....9A,2013EPJC...73.2364B}.
%Measurements of antiproton spectra in $pp$- and $p\,{\rm Be}$-collisions 
%at 19.2 GeV/c \citep{allaby70,1975NuPhB..86..403A} and in $pp$- and $p\,{\rm C}$-interactions 
%at 158 GeV/c \citep{2010EPJC...65....9A,2013EPJC...73.2364B}. 
In addition, to demonstrate
that the introduced modifications have not spoiled the treatment
of high energy interactions, we compare in Fig.~\ref{fig:ap900}
the calculated transverse momentum spectrum of antiprotons in $pp$-collisions
at $\sqrt s=900$\,GeV with the data of the ALICE experiment~\citep{2011EPJC...71.1655A}.
As can be judged from the Figures, there is a reasonable overall description of $\bar{p}$-production
over a wide energy range. 
%has been achieved. 
The differences with the results 
of the EPOS-LHC model can be used as a measure for model uncertainties 
in the calculations of secondary antiproton spectrum using QCD MC generators.
For comparison, we plot in  Figs.~\ref{fig:ap158} and \ref{fig:ap900} 
also the results obtained using the original QGSJET-II-04 model. At 
$\sqrt s =900$\,GeV, the differences between the models QGSJET-IIm
and QGSJET-II-04 are pretty small. On the other hand,
at lower energies there is a significant reduction
of the antiproton yield predicted by QGSJET-IIm, which reaches 
$\simeq 20$\% already at 158\,GeV/c.

%%%%%%%%%%%%%%%%%%%%%%%%%%%%%%%%%%%%%%%%%%%%%%%%%%%%%%%%%%%%%%%%%%%%%%%%%%%%%%%
\section{$Z-$factors for $\bar{p}$ production: comparison of model 
predictions} \label{sec:zfactors}

To compare the impact of different interaction models and
parameterizations on the predicted CR antiproton spectrum, it is convenient,
similarly to the $\gamma$-ray case \citep{2014ApJ...789..136K}, to use the corresponding
``$Z$-factors.'' They are defined as the spectrally averaged energy fraction 
transferred to antiprotons in proton-proton, proton-nucleus, nucleus-proton, or
nucleus-nucleus collisions, assuming that the
spectra of CR species in the relevant energy range can be approximated
by a power-law, $I_i(E)=K_i\, E^{-\alpha_i}$.
Then the contribution $q^{ij}_{\bar p}(E_{\bar p})$ to the flux of secondary
CRs (here, antiprotons) from interactions of the CR species
$i$ with ISM component $j$ ($i,j=$ protons or nuclei) of number density $n_j$,
\be
\label{qij}
 q^{ij}_{\bar p}(E_{\bar p}) = n_j \int_{E_{\rm thr}(E_{\bar p})}^\infty d E \;
 \frac{d\sigma^{ij\rightarrow \bar p}(E,E_{\bar p})}
		 {d E_{\bar p}} \: I_i(E),
\ee
can be rewritten as~\citep{2014ApJ...789..136K}
\be 
\label{q_ij-Z}
 q^{ij}_{\bar p}(E_{\bar p}) = n_j \, I_i(E_{\bar p})
 \, Z^{ij}_{\bar p}(E_{\bar p},\alpha_i)\,.
\ee
Here, we expressed the $Z$-factor $Z^{ij}_{\bar p}$ via the inclusive
spectra of antiprotons $d\sigma^{ij\rightarrow \bar p}(E,z_{\bar p})/d z_{\bar p}$, 
$z_{\bar p}=E_{\bar p}/E$, as
\be 
\label{Z_spec}
 Z^{ij}_{\bar p}(E_{\bar p},\alpha) = \int_0^1 d z \, z^{\alpha-1}\,
 \frac{d\sigma^{ij\rightarrow \bar p}(E_{\bar p}/z,z)}{d z}\,. 
\ee
Note that $E$ corresponds to the energy per nucleon for nuclear projectiles.

The $Z$-factors $Z^{ij}_{\bar p}$ depend clearly both on the $\bar{p}$ 
production spectra and on the spectral slopes $\alpha_{i}$, containing 
all the dependences on hadronic interaction models. On the
other hand, these factors are independent of the CR abundances.
Using two different interaction models, M1 and M2, 
the ratio of the respective contributions to the
secondary fluxes equals the  ratio of the corresponding $Z$-factors
[c.f.\ Eq.\ (\ref{q_ij-Z})]:
\be
\frac{q^{ij}_{\bar p{\rm (M1)}}(E_{\bar p})}
{q^{ij}_{\bar p{\rm (M2)}}(E_{\bar p})}
=\frac{Z^{ij}_{\bar p{\rm (M1)}}(E_{\bar p},\alpha_i)}
{Z^{ij}_{\bar p{\rm (M2)}}(E_{\bar p},\alpha_i)}\,.
\ee

\begin{figure}[p]
\center{
\includegraphics[width=0.45\textwidth,height=6cm]{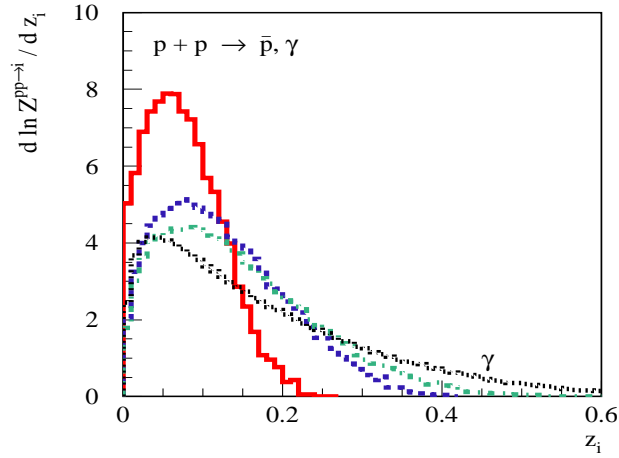}
}
\caption{Distribution of spectrum-weighted ($\alpha =2.6$)
energy fraction $z_i$ ($i=\bar p, \gamma$). For antiprotons, the
distribution of
$z_{\bar p}=E_{\bar p}/E$ is shown for
different energies 
$E_{\bar p}^{\rm kin}=$ 1\,GeV
(solid, red), 3\,GeV (dashed, blue), and 10\,GeV (dot-dashed, green).
A similar distribution for  $z_{\gamma}=E_{\gamma}/E$ 
for the case of $\gamma$-ray
production is shown for $E_{\gamma}=10$\,GeV by the dotted, black line marked ``$\gamma$''.
\label{fig:z_pbar}}
\end{figure}

In the following, we are going to compare the factors $Z^{pp}_{\bar{p}}$
obtained with the modified QGSJET-IIm model and with EPOS-LHC to the
$Z$-factors calculated using some widely used parameterizations of $\bar{p}$-spectra
for $pp$-collisions. However, before doing so, let us investigate
which projectile energies contribute mainly to $Z^{pp}_{\bar{p}}(E_{\bar{p}},\alpha)$.
To this end, we plot in Fig.~\ref{fig:z_pbar} the spectrally-weighted
(for definiteness, we use $\alpha =2.6$)
distribution of the energy fraction $z_{\bar p}=E_{\bar p}/E$ transferred
to antiprotons  
\be
\frac{1}{Z^{pp}_{\bar p}}\,\frac{dZ^{pp}_{\bar p}}{dz_{\bar p}}
=z_{\bar p}^{\alpha-1}\,
 \frac{d\sigma^{ij\rightarrow \bar p}
 (E_{\bar p}/z_{\bar p},z_{\bar p})}{d z_{\bar p}}\label{dZ/dz}
\ee
for different kinetic energies $E_{\bar p}^{\rm kin}=E_{\bar p}-m_{\bar p}$.
For comparison, the same distribution for $\gamma$-ray production,
$1/Z^{pp}_{\gamma}\times dZ^{pp}_{\gamma}/dz_{\gamma}$
($z_{\gamma}=E_{\gamma}/E$), is also shown.
Clearly, the range of CR proton energies $E=E_{\bar p}/z_{\bar p}$
which contributes significantly 
to antiproton production is substantially narrower compared to the
$\gamma$-ray case.
Since the $\bar p$ spectrum is much softer than   that  of gammas, 
the region of moderately large $z$ contributes much less
to the production of $\bar p$'s than to the production of $\gamma$'s. Additionally, the distribution
becomes substantially narrower with decreasing $E_{\bar{p}}$, which is
a consequence of both threshold effects and
the interaction kinematics: For small $E_{\bar p}^{\rm kin}$,
the region of not too small $z_{\bar p}$ becomes kinematically
forbidden,
because the corresponding proton energy $E=E_{\bar p}/z_{\bar p}$
falls below the production threshold. On the other hand, the contribution
of the region $z_{\bar p}\rightarrow 0$ is strongly suppressed by the 
spectral factor $z_{\bar p}^{\alpha-1}$, c.f.~Eq.~(\ref{dZ/dz}).
Moreover, for a given CR proton energy
$E$, antiprotons are produced most copiously in the c.m.s.\
central region ($x_{\rm F}\sim 0$), which corresponds to 
$E_{\bar{p}}\sim \sqrt{2m_N E}$ in the lab.\ frame (where $m_N$ is the
nucleon mass). Thus, the region $E_{\bar{p}}\ll \sqrt{2m_N E}$ 
or $z_{\bar p}\ll \sqrt{2m_N/E}$ corresponds to
the target fragmentation region in the c.m.s., 
while $E_{\bar p}^{\rm kin}\rightarrow 0$ 
  corresponds to  the kinematic 
boundary ($x_{\rm F}\rightarrow -1$ in c.m.s.), where
the $\bar p$ spectrum falls down rapidly. 
This can be clearly seen in Fig.~\ref{fig:ap158}~(right) for the case
of proton-carbon collisions.

The calculated energy dependence of the $Z$-factors for $\bar p$
production\footnote{Here and in the following we take into account both
$\bar p$ and $\bar n$ production when calculating $Z$-factors;
for brevity, we use the same notation $Z^{pp}_{\bar{p}}$ instead of
$Z^{pp}_{\bar{p}+\bar{n}}$. For the parameterizations by
\citet{1983JPhG....9.1289T} and \citet{2003PhRvD..68i4017D}, we simply double the respective $\bar p$ yields.},
$Z^{pp}_{\bar{p}}(E_{\bar{p}},\alpha)$, is compared in Fig.~\ref{fig:zfac-pp}
for QGSJET-IIm, EPOS-LHC, and the parameterizations from
\citet{1983JPhG....9.1289T} and \citet{2003PhRvD..68i4017D}. We consider two values, $\alpha =2$ and $\alpha =3$, 
for the slope of the CR proton spectrum, which bracket the physically
most interesting range.

\begin{figure}[p]
\center{
\includegraphics[width=0.45\textwidth]{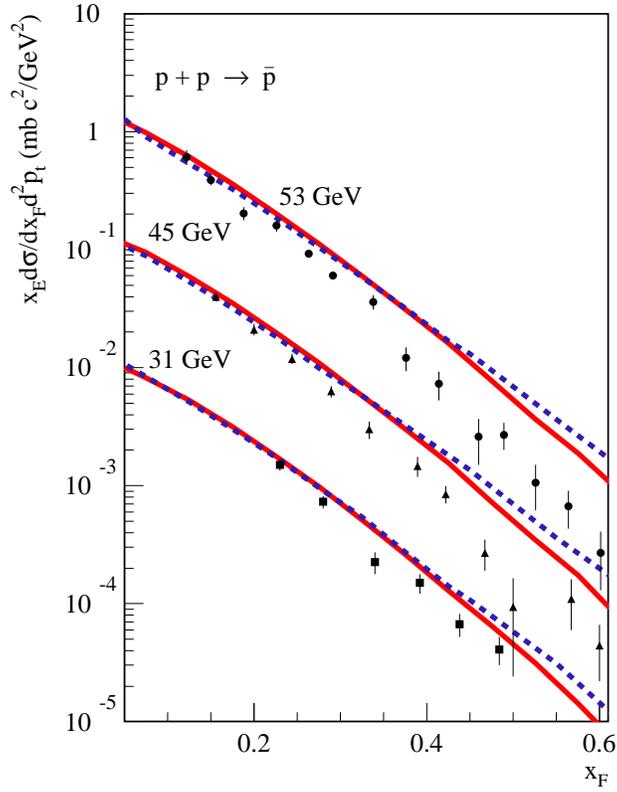}
}
\caption{Feynman $x$-spectra of $\bar p$'s 
for fixed c.m.s.\ angle $\tan \theta = 2.66/\sqrt{s}$ in $pp$-collisions at $\sqrt{s}=31$ ($\times 10^{-2}$), 45 ($\times 10^{-1}$), and 53\,GeV,
calculated using QGSJET-IIm (solid, red) and EPOS-LHC (dashed, blue), in comparison with
experimental data \citep{1973NuPhB..56..333A}. 
\label{fig:ap-sps-epos}}
\end{figure}

Comparing first the results of QGSJET-IIm and EPOS-LHC, we observe a 
relatively good agreement between them in the high energy range
for $\alpha =2$, while somewhat larger differences 
are obtained for a steeper slope ($\alpha =3$) and for 
$E_{\bar p}^{\rm kin}\lesssim 100$\,GeV. This is due to
the harder $\bar p$ production spectra predicted by EPOS-LHC
(c.f.~Figs.~\ref{fig:ap19}, \ref{fig:ap158}, and \ref{fig:ap-sps-epos}).
However, the results for the two MC generators show a
reasonable overall agreement and, as already stated, the remaining 
differences can be used as a measure for the model uncertainties.

Next we consider differences between the modified QGSJET-IIm model and
the parameterized $\bar p$ spectra from \citet{1983JPhG....9.1289T} and \citet{2003PhRvD..68i4017D}.
In the energy range $E_{\bar p}^{\rm kin}=10$--100\,GeV, where the 
relevant proton-proton interactions are covered by fixed target experiments,
the $\bar p$ spectra calculated with these parameterizations
agree approximately with those obtained using QGSJET-IIm.
However, at higher energies 
the parameterized results of \citet{1983JPhG....9.1289T} and \citet{2003PhRvD..68i4017D} are rather 
unreliable, which is best illustrated by the large difference between the
$Z$-factors calculated using the two parameter sets proposed 
by~\citet{2003PhRvD..68i4017D}.
Not surprisingly, the differences increase for larger $\alpha$
due to the stronger sensitivity to the forward $\bar p$-production spectra which are less constrained by experimental data.

To get a better  idea of
these differences, we plot in 
Fig.~\ref{fig:zdiff} the ratio of $Z$-factors $Z^{pp}_{\bar{p}}$ calculated
with QGSJET-IIm and with the three parameterizations considered. One
immediately notices a sizeable enhancement of the $\bar p$ yield
in the former case for $E_{\bar p}^{\rm kin}\lesssim 10$\,GeV.
The differences with the results obtained using
the parameterizations of \citet{1983JPhG....9.1289T} and
 \citet{2003PhRvD..68i4017D} reach a factor of two
for $E_{\bar p}^{\rm kin}\simeq 1$\,GeV and originate from the 
treatment of $\bar p$ production close to the kinematic threshold
(c.f.~Fig.~\ref{fig:z_pbar}), which is rather weakly constrained by
available experimental data. Thus, these differences may be regarded
as characteristic uncertainties for calculations of the $\bar p$-yield
in this low energy range.

More importantly, for $E_{\bar p}^{\rm kin}\gtrsim 100$\,GeV, 
the results obtained using 
the parameterizations of \citet{1983JPhG....9.1289T} and \citet{2003PhRvD..68i4017D} start to diverge 
strongly from the results based on QGSJET-IIm, as can be clearly seen in 
Figs.~\ref{fig:zfac-pp} and \ref{fig:zdiff}. For instance,
using the $\bar p$-production spectra from \citet{1983JPhG....9.1289T}, which is the
standard reference for astrophysical applications, one observes that
the respective $Z^{pp}_{\bar{p}}$
becomes practically energy-independent for
$E_{\bar p}^{\rm kin}\gtrsim 100$\,GeV
(see Fig.~\ref{fig:zfac-pp}), which is a consequence of the scaling
picture used in their ansatz. However, the assumed scaling behavior for
the inclusive particle production spectra 
is explicitly broken by the energy rise of the inelastic $pp$ cross section 
$\sigma^{\rm inel}_{pp}$, which has been firmly established by numerous
accelerator experiments from fixed target experiments to those at the LHC.
Moreover, the $Z$-factors for $\bar p$-production are rather sensitive to 
the central region in the c.m.s., because the $\bar p$-production spectrum in
$pp$-collisions is soft. In this region, the particle density rises quicker 
than $\sigma^{\rm inel}_{pp}$, namely, as a power law \citep[see, e.g., the 
discussion in][]{2011APh....35...98D}. The resulting
enhancement of $\bar p$-production by high-energy CRs, as shown by
the ratio of the respective $Z$-factors, is quite significant and may
have an important impact on the interpretations of 
all sorts of CR data, not only $\bar p$, by the
PAMELA and AMS2 experiments in this energy region
\citep[see, e.g.,][]{
2003A&A...410..189B,
2009PhRvL.103e1104B,
2009PhRvL.103h1103B,
2009JCAP...01..025H,
2011ARA&A..49..155P,
2011ApJ...733..119K,
2013PhRvD..87d7301K,
2014JCAP...04..006D,
2014PhRvD..89d3013C,
2014PhRvD..90f1301M}.

Both parameterizations proposed by \citet{2003PhRvD..68i4017D} are based on the 
erroneous ansatz of \citet{1983ApJS...51..271L} for $\sigma^{\rm inel}_{pp}$ and 
$\sigma^{\rm inel}_{pA}$, i.e.\ they assume that the inelastic cross section 
is constant for $E_{p}^{\rm kin}\gtrsim 1$\,GeV. The additional assumptions
made by \citet{2003PhRvD..68i4017D} about the energy dependence of the very forward 
part of $\bar p$-production spectra, which are responsible for the huge
variations between the two parameterizations (Fig.~\ref{fig:zfac-pp}),
are also questionable as we discuss in more detail in Appendix~\ref{Ap-B}.

%%%%%%%%%%%%%%%%%%%%%%%%%%%%%%%%%%%%%%%%%%%%%%%%%%%%%%%%%%%%%%%%%%%%%%%%%%%%
\section{Nuclear enhancement} \label{sec:enh}

Now we proceed with calculating the nuclear enhancement of the $\bar p$-yield
due to contributions of nuclear species in CRs and of the
helium component in the ISM. As follows from Eq.~(\ref{q_ij-Z}) and the
more detailed discussion by~\citet{2014ApJ...789..136K}, the respective partial 
enhancements compared to the yield from $pp$-interactions are proportional 
to the corresponding $Z$-factors,
 \be \label{eq:qenh}
 \epsilon _{ij}^{\bar p}(E_{\bar p}) = \frac{q^{ij}_{\bar p}(E_{\bar p})}
 {q^{pp}_{\bar p}(E_{\bar p})}=\frac{n_j}{n_p} \:
\frac{I_i(E_{\bar p})}{I_p(E_{\bar p})} \:
\frac{Z^{ij}_{\bar p}(E_{\bar p},\alpha_i)}
{Z^{pp}_{\bar p}(E_{\bar p},\alpha_p)}\,.
\ee
In particular, contributions of CR nuclei may be additionally enhanced,
if the spectral indices  $\alpha _i<\alpha _p$, because of the strong $\alpha$-dependence of 
the $Z$-factors \citep{2014ApJ...789..136K}.

\begin{figure}[p]
\center{
\includegraphics[width=0.45\textwidth]{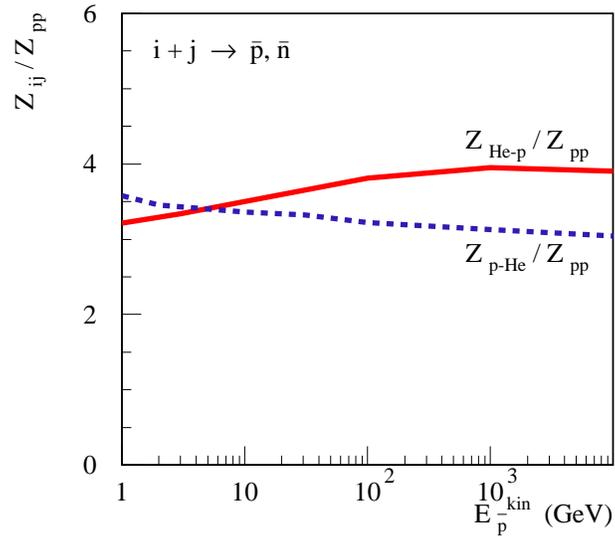}
}
\caption{Energy dependence of the enhancement of the He\,$p$ (solid, red) 
and $p$\,He (dashed, blue) contributions to the 
antiproton spectrum, relative to the $pp$-case,
 $Z^{ij}_{\bar p}(E_{\bar p},\alpha)/Z^{pp}_{\bar p}(E_{\bar p},\alpha)$ 
(plotted as a function of $E_{\bar p}^{\rm kin}$),
for $\alpha =2.6$.\label{fig:enh}}
\end{figure}

As follows from the general analysis \citep{2014ApJ...789..136K}, in the limit of high
energies and large $\alpha$, one expects the following simple relations to 
hold,
\begin{eqnarray}
\frac{Z^{ij}_{\bar p}(E_{\bar p},\alpha)}
{Z^{pj}_{\bar p}(E_{\bar p},\alpha)}& \simeq &i, \label{eq:superp}\\
\frac{Z^{ij}_{\bar p}(E_{\bar p},\alpha)}
{Z^{ip}_{\bar p}(E_{\bar p},\alpha)}& \simeq &
 \frac{\sigma^{\rm inel}_{ij}}{\sigma^{\rm inel}_{ip}}<j\,. \label{eq:scal}
 \end{eqnarray}
However, as illustrated in Fig.~\ref{fig:enh} for the case of 
He-$p$ and $p$-He collisions,
at finite energies and for realistic spectral slopes these relations
are strongly modified by threshold effects related to the relatively
large antiproton mass. Indeed, the ratio 
$Z^{{\rm He}\,p}_{\bar p}/Z^{pp}_{\bar p}$ approaches the asymptotic value 4
very slowly. Due to the soft $\bar p$-spectrum, up to rather high energies,
$Z^{Ap}_{\bar p}$ remains sensitive to $\bar p$-production in the backward
c.m.s.\ region where the discussed ``$A$-enhancement'' does not hold.
The same effect is partly responsible for the rise of the ratio
$Z^{p\,{\rm He}}_{\bar p}/Z^{pp}_{\bar p}$ for decreasing $E_{\bar p}$ as can be
seen in Fig.~\ref{fig:enh}.
As discussed in the previous Section, for small $E_{\bar p}^{\rm kin}$
the contribution of the forward c.m.s.\ region (moderately large 
$z_{\bar p}$) is substantially reduced by threshold effects.
As a consequence, the relative contribution of the backward c.m.s.\ 
region $z_{\bar p}\ll \sqrt{2m_N/E_p}$ is increased, resulting
in a ``$A$-enhancement'' of the $\bar p$-yield in $pA$-collisions. 
As the bottom line, we conclude that the nuclear enhancement of $\bar p$-production is strongly energy dependent in the kinematic range
relevant for astrophysical applications. Hence, the use of a constant
``nuclear enhancement factor'' for calculations of CR antiproton spectra
is difficult to justify. 
 
In Table~\ref{tab:z-factors}, 
we collect the calculated $Z$-factors
$Z^{ij}_{\bar p}$ for different $E_{\bar p}^{\rm kin}$, slopes $\alpha$,
and different combinations of CR and ISM nuclei.
These results may be used for the
calculation of secondary antiproton spectra when the
partial spectra of CR mass groups can be approximated by a
power law behavior in the corresponding energy range.
\emph{As an illustration}, we have calculated the energy dependence of the
nuclear enhancement factor for $\bar p$ production, 
$\epsilon_{\bar p}(E_{\bar p})=\sum_{i,j}\epsilon _{ij}^{\bar p}(E_{\bar p})$,
with $\epsilon _{ij}^{\bar p}$ defined in Eq.\ (\ref{eq:qenh}),
in the energy range $E_{\bar p}^{\rm kin}=10$\,GeV--10\,TeV, using
the high energy limit of the parameterization for the spectra of CR species
by \citet{2004PhRvD..70d3008H}; 
the respective parameters $K_i$ and $\alpha_i$ are given 
in Table \ref{tab:para} for convenience.
In contrast to all previous calculations, our results for $\epsilon_{\bar p}$,
presented in Table \ref{tab:enh-factor}, indicate a strong energy rise of the
nuclear enhancement for secondary antiprotons, which reaches a factor of
two for $E_{\bar p}^{\rm kin}\simeq 1$ TeV.  
\emph{For this particular parameterization of the spectra of CR species
by \citet{2004PhRvD..70d3008H},} the energy dependence of the 
nuclear enhancement factor $\epsilon_{\bar p}$ can be described by a
power-law, $\epsilon_{\bar p}\approx 1.58\times (E/{\rm GeV})^{0.034}$.

\begin{figure}[p]
\center{
\includegraphics[width=0.45\textwidth]{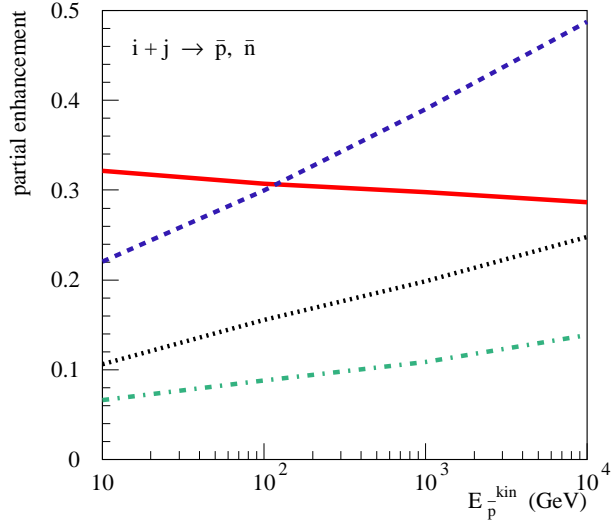}
}
\caption{Energy dependence of partial contributions 
 $\epsilon _{ij}^{\bar p}(E_{\bar p})$   to the nuclear enhancement
factor from different
interaction channels:  $p$\,He (solid, red), He\,$p$ (dashed, blue),
He\,He (dot-dashed, green), and all others (dotted, black); the CR composition
given in Table~\ref{tab:para} is used.
\label{fig:enh-part}}
\end{figure}

To understand better this result,
we plot in Fig.~\ref{fig:enh-part} the partial contributions 
 $\epsilon _{ij}^{\bar p}(E_{\bar p})$ to the nuclear enhancement
factor from different interaction channels. One immediately notices
a steep energy rise of the antiproton yield from interactions of CR
helium with ISM protons. For the CR composition considered,
the reason for this rise is three-fold. Apart from the trivial increase
of the fraction of CR helium, the flatter helium spectrum compared
to protons ($\alpha _{\rm He}<\alpha _p$) enhances the rise of 
$\epsilon _{{\rm He}-p}^{\bar p}(E_{\bar p})$ because of   the strong
$\alpha$-dependence of the $Z$-factors $Z^{ij}_{\bar p}(E_{\bar p},\alpha)$
(c.f.\ Eq.\ (\ref{eq:qenh}) and Table  \ref{tab:para}), as noticed previously in
\citet{2014ApJ...789..136K} for $\gamma$-ray production. 
%%%
Finally, the energy dependence of the $Z$-factors is affected by 
threshold effects, as we have already noticed in Fig.~\ref{fig:enh}.
As a result, the partial contributions of CR nuclei
in the region of relatively small $E_{\bar p}$ are suppressed.
This suppression diminishes at higher energies, where one approaches the 
asymptotic limit of Eq.~(\ref{eq:superp}). The latter effect  
dominates the energy rise of the contribution of heavier nuclei
(CNO, Mg-Si, and Fe), which is shown in Fig.~\ref{fig:enh-part} 
by the dotted line. In contrast to expectations, their contribution %of heavier nuclei 
is significant in the TeV range: The combined
antiproton yield from interactions of the heavier nuclei with ISM protons 
and helium becomes at high energies comparable
in magnitude with the one from the proton-helium channel.
%%%%
%The third effect is
%  the above-discussed influence of threshold effects on the energy dependence
%  of the  $Z$-factors (c.f.\ Fig.\ \ref{fig:enh}), 
%  which substantially suppress partail contributions of primary nuclei
%  in the region of relatively small $E_{\bar p}$. This suppression diminishes
%  when moving to higher energies and one approaches the asymptotic limit
%  of Eq.\ (\ref{eq:superp}). In particular, it is the latter effect which 
%  dominates the energy rise of the contributions of heavier nuclei
%  (CNO, Mg-Si, and Fe), shown if the Figure by the dotted line.
%  In contrast to   common expectations, the latter appear to be very
%  significant in the TeV range: the summary contribution of interactions
%  of the heavier nuclei with ISM protons and helium becomes comparable
%  in magnitude with the one from the proton-helium channel.

%%%%%%%%%%%%%%%%%%%%%%%%%%%%%%%%%%%%%%%%%%%%%%%%%%%%%%%%%%%%%%%%%%%%%%%%%%%%%%
\section{Conclusions} \label{sec:conclusion}

Accurate antiproton production cross sections are a critical pre-requisite
for many astrophysical applications and searches for new physics. 
We have used, therefore, EPOS-LHC and QGSJET-II-04, two of the most 
advanced Monte Carlo generators which reproduce numerous accelerator data
including the most recent ones from LHC, to calculate the antiproton 
yield in $pp$-interactions.  In the case of QGSJET-II-04,
a tune of its fragmentation procedure was required for an adequate 
description of particle production at low energies and high thresholds.
After that, we have found that the antiproton yields
of the two QCD  Monte Carlo generators agree reasonably well with each other
and the available experimental data. Therefore, we conclude that the results 
of these two generators can be used to predict  reliably the antiproton 
yield outside the energy range covered by fixed target accelerator data, 
$E_{\bar p}\approx 10-$100\,GeV. 
Moreover, using these Monte Carlo generators it is straightforward
to calculate also the antiproton yield in $pA$- and $AA$-interactions.
In the limiting case, when the spectra of CR species  can be approximated by a power-law,
we have derived the nuclear enhancement of the $\bar p$-yield
due to contributions of nuclear species in CRs and of the
helium component in the ISM. In contrast to all previous calculations, 
our results indicate a strong rise of the
nuclear enhancement for secondary antiprotons   with energy, which reaches a factor of
two for $E_{\bar p}^{\rm kin}\simeq 1$ TeV.

We have also compared our results obtained using  EPOS-LHC and QGSJET-II-04
to the commonly used parameterizations of the antiproton yield from 
\citet{1983JPhG....9.1289T} and \citet{2003PhRvD..68i4017D}. 
In the energy range $E_{\bar p}^{\rm kin}=10$--100\,GeV, where the 
relevant proton-proton interactions are covered by fixed target experiments,
these parameterizations agree approximately with the results obtained using 
QGSJET-IIm. At higher energies, when these parameterizations are extrapolated
outside the kinematic range constrained by experimental data, the
differences increase fast, because the physical   concepts  used in selecting
their fitting functions are incorrect. In particular, the assumed scaling behavior 
for the inclusive production spectra is broken by the energy rise of 
$\sigma_{pp}^{\rm inel}$, invalidating their ansatz. The resulting increase
of the $\bar p$ production should be taken into account in the
interpretation of CR data, especially those at $E>100$\,GeV
 expected from AMS-02, ISS-CREAM, and CALET experiments.  

In the   low energy domain  $E_{\bar p}^{\rm kin}\lsim 10$\,GeV, the differences between
the results obtained using  EPOS-LHC and QGSJET-II-04 and the
parameterizations of \citet{1983JPhG....9.1289T} and
\citet{2003PhRvD..68i4017D}   are also significant, reaching a factor of two
at $E_{\bar p}^{\rm kin}\simeq 1$\,GeV (c.f.\ Fig.~\ref{fig:zdiff}). 
The origin of these differences
can be traced to the treatment of $\bar p$-production close to the 
kinematic threshold, which is rather weakly constrained by
available experimental data. Since this energy range is the most
interesting for dark matter searches, additional experimental data
from, e.g., the NA61 experiment are highly desirable. In the
absence of these data, the uncertainty of the antiproton yield
in the low-energy range should be increased, reaching $\sim 50\%$
at $E_{\bar p}^{\rm kin}=1$\,GeV.

\acknowledgements

The authors are grateful to Tanguy Pierog for useful discussions
and for his help in
comparing predictions of various Monte Carlo generators to accelerator
data on antiproton production.
IVM and SSO acknowledge support from NASA through the grants 
NNX13AC47G and NNX13A092G.

\vspace*{5mm}
\noindent
Addendum: When this paper was already prepared for a submission,
new parameterizations for $\bar p$-production in $pp$-collisions
have been proposed in two recent publications \citep{2014PhRvD..90h5017D,2014JCAP...09..051K}.
\citet{2014PhRvD..90h5017D} used essentially the same parameterization
as \citet{2003PhRvD..68i4017D} and also assumed that the $\bar p$-production 
spectrum scales as a power of energy ($\propto s^{\Delta}$) in the high 
energy limit. One of their proposed parameterizations 
cannot be extrapolated beyond the domain of validity which depends
both on the energy and the slope of the primary spectra. In particular,
the parametrization is not suited for a power-law primary energy spectrum
$I_p(E)\sim E^{-\alpha_p}$ with $\alpha_p<2.7$.
%
%is ill-behaved outside the fitting range, producing meaningless 
%results for the antiproton yield (e.g.\ the yield becomes infinitely large,
%if the slope $\alpha_p$ of the energy spectrum of CR proton becomes
%$\alpha_p<2.75$).  Moreover, the numerical results presented 
%in~\citet{2014arXiv1408.0288D1} seem to be unreliable.
In turn, \citet{2014JCAP...09..051K}
followed the approach of \citet{1983JPhG....9.1289T} by using the outdated
concept of radial scaling, which is broken, e.g., by the energy rise
of the inelastic cross section, as already discussed above.

\appendix
%%%%%%%%%%%%%%%%%%%%%%%%%%%%%%%%%%%%%%%%%%%%%%%%%%%%%%%%%%%%%%%%%%%%%%%%%%%%%%
\section{Comparison with differential $\bar p$ spectra} \label{Ap-A}

\begin{figure}[p]
\center{
\includegraphics[width=0.45\textwidth]{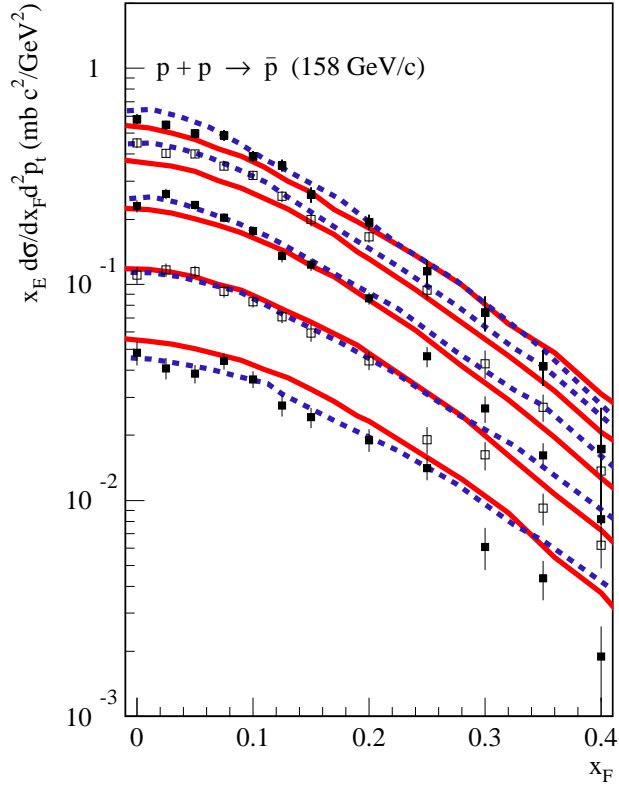}
}
\caption{Feynman $x$-spectra of $\bar p$'s (c.m.s.) 
for fixed $p_{\rm t}$ (from top to bottom:
0.1, 0.3, 0.5, 0.7, and 0.9\,GeV) in $pp$-collisions at $p_{\rm lab}=158$\,GeV/c, 
calculated using QGSJET-IIm (solid, red) and EPOS-LHC (dashed, blue), in comparison with NA49
data \citep{2010EPJC...65....9A}.\label{fig:ap158-diff}}
\end{figure}

In addition to the $p_{\rm t}$-integrated spectra of $\bar p$'s
for $pp$-collisions at 158\,GeV/c, plotted in Fig.~\ref{fig:ap158},
we present in Fig.~\ref{fig:ap158-diff} the results of QGSJET-IIm and EPOS-LHC
for the respective differential spectra 
$x_E\,d\sigma /dx_{\rm F}/d^2p_{\rm t}$ for fixed values of $p_{\rm t}$.
Comparing the calculations to NA49 data \citep{2010EPJC...65....9A}, we observe again
a relatively good agreement, though the calculated spectra are somewhat
harder than the measured ones.

The same tendency is indicated by a comparison with data from the
CERN Intersecting Storage Rings (ISR)~\citep{1973NuPhB..56..333A}
at higher energies, see Fig.~\ref{fig:ap-sps-epos}, 
though any definite conclusions are hampered by both the large systematic
errors of the measurements and by the narrow kinematic coverage of the 
experiment\footnote{The measurement has been performed with the
spectrometer technique for a single fixed c.m.s.\ angle, thus covering a tiny
fraction of the relevant kinematic space.}.
 
\begin{figure}[p]
\center{
\includegraphics[width=0.45\textwidth]{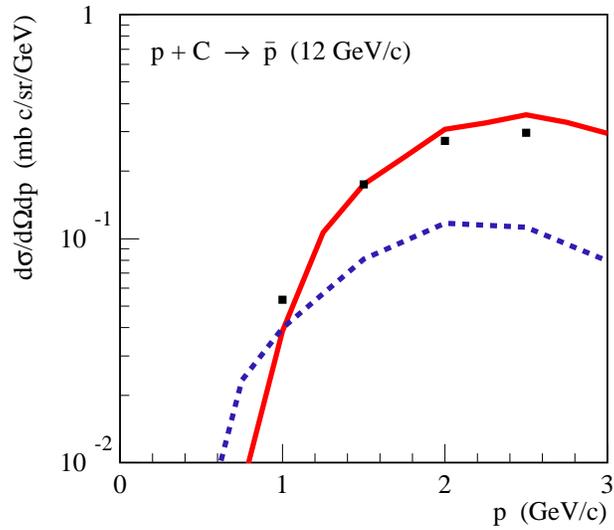}
}
\caption{Momentum spectra of $\bar p$'s in the laboratory frame 
for $\theta =5.1^\circ$   in $p$\,C-collisions  at $p_{\rm lab}=12$\,GeV/c,
calculated using QGSJET-IIm (solid, red) and EPOS-LHC (dashed, blue),
in comparison with experimental data \citep{1998NuPhA.634..115S}.
\label{fig:ap12-epos}}
\end{figure}

Finally, the results of the two models for the momentum spectrum of $\bar p$'s
in $p$\,C-collisions at 12\,GeV/c are compared in 
Fig.~\ref{fig:ap12-epos} to spectrometer measurements by
\citet{1998NuPhA.634..115S}.
The good agreement of the spectrum calculated with QGSJET-IIm
with the data may be regarded as an indication that the results
of the model remain reasonable even   when  approaching the kinematic threshold
for $\bar p$-production. However, again, the narrow kinematic coverage
of the experiment does not allow one to make any definite conclusions.

%%%%%%%%%%%%%%%%%%%%%%%%%%%%%%%%%%%%%%%%%%%%%%%%%%%%%%%%%%%%%%%%%%%%%%%
\section{Comparison of parameterizations of $\bar p$ spectra with 
accelerator data} \label{Ap-B}

\begin{figure*}[p]
\center{
\includegraphics[width=0.9\textwidth]{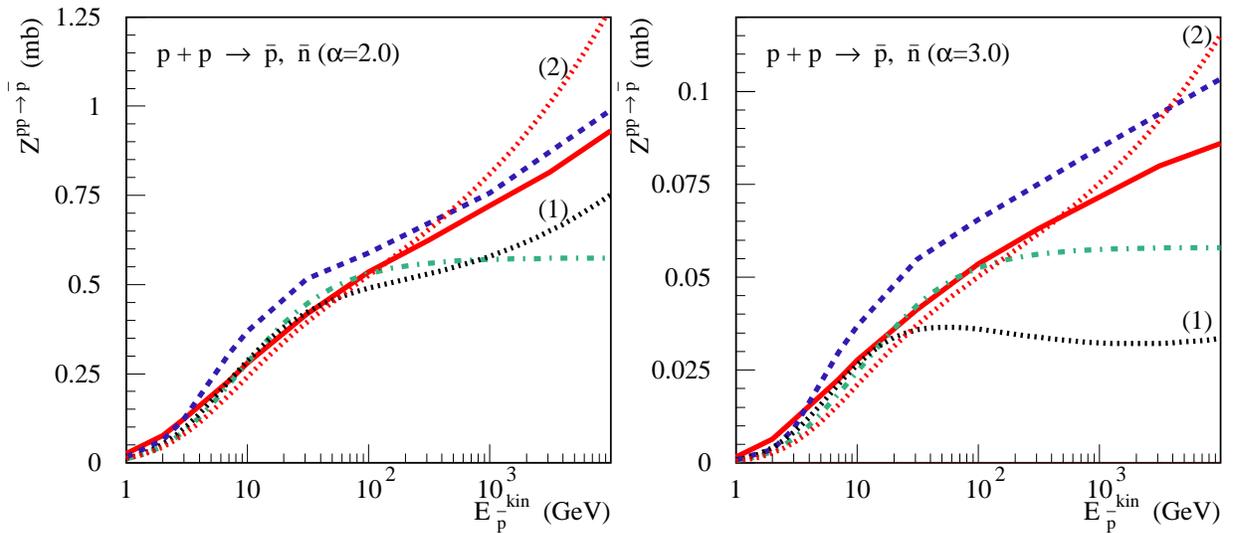}
}
\vskip5mm
\caption{Energy dependence of $Z$-factors   for $\bar p$ and $\bar n$
 production,
 $Z^{pp}_{\bar{p}}(E_{\bar{p}},\alpha)$ (plotted as a function of
$E_{\bar p}^{\rm kin}$), for 
$\alpha=2$ (left) and $\alpha=3$ (right), as calculated with 
QGSJET-IIm (solid, red) and EPOS-LHC (dashed, blue), or using the parameterizations
by \citet{1983JPhG....9.1289T} (dot-dashed, green) and \citet{2003PhRvD..68i4017D} (dotted; the 
lines marked ``(1)'' (black) and ``(2)'' (red) correspond to the respective parameter sets). 
\label{fig:zfac-pp}} 
\end{figure*}

\begin{figure}[p]
\center{
\includegraphics[width=0.45\textwidth]{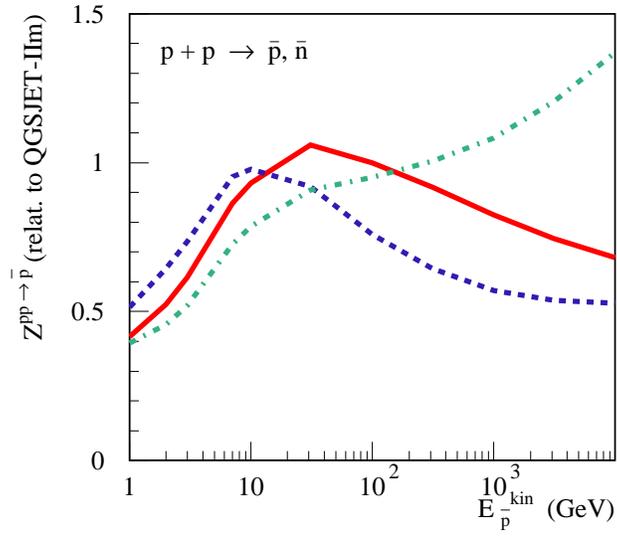}
}
%\caption{Ratios of $Z$-factors for ($\bar p + \bar n$) production, $Z^{pp}_{\bar p}$,
%obtained using parameterizations of $\bar p$-spectra:
%\citet{1983JPhG....9.1289T} - solid, set 1 of \citet{2003PhRvD..68i4017D} - dashed,
%set 2 of \citet{2003PhRvD..68i4017D} - dot-dashed, to the one calculated with QGSJET-IIm,
%for $\alpha =2.6$.
\caption{Ratios of $Z$-factors    for $\bar p$ and $\bar n$ production, 
$Z^{pp}_{\bar p}$,
obtained using various parameterizations of $\bar p$-spectra to 
the $Z$-factor calculated with QGSJET-IIm,
for $\alpha =2.6$:
solid red -- \citet{1983JPhG....9.1289T}, 
dashed blue -- set 1 of \citet{2003PhRvD..68i4017D},
dot-dashed green -- set 2 of \citet{2003PhRvD..68i4017D}. 
\label{fig:zdiff}}%
\end{figure}

\begin{figure}[p]
\center{
\includegraphics[width=0.45\textwidth]{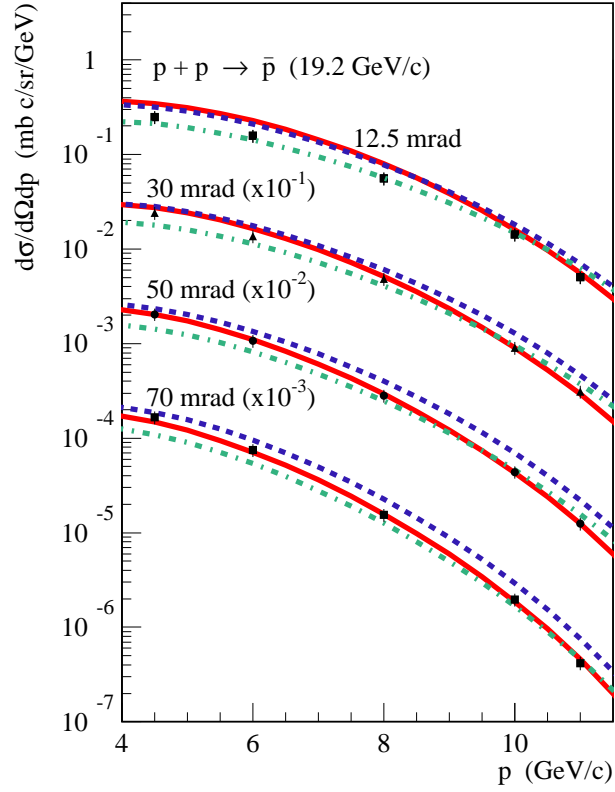}
}
\caption{Momentum spectra of $\bar p$'s in the laboratory frame 
for different laboratory angles $\theta$ (as indicated in the plots)
in $pp$-collisions at $p_{\rm lab}=19.2$\,GeV/c, calculated using parameterizations by \citet{1983JPhG....9.1289T} (solid, red)
or by \citet{2003PhRvD..68i4017D} (parameter set 1 - dashed, blue, parameter set 2 - dot-dashed, green), in comparison with experimental data 
\citep{allaby70,1975NuPhB..86..403A}.\label{fig:ap19-ng}}
\end{figure}

\begin{figure}[p]
\center{
\includegraphics[width=0.4\textwidth]{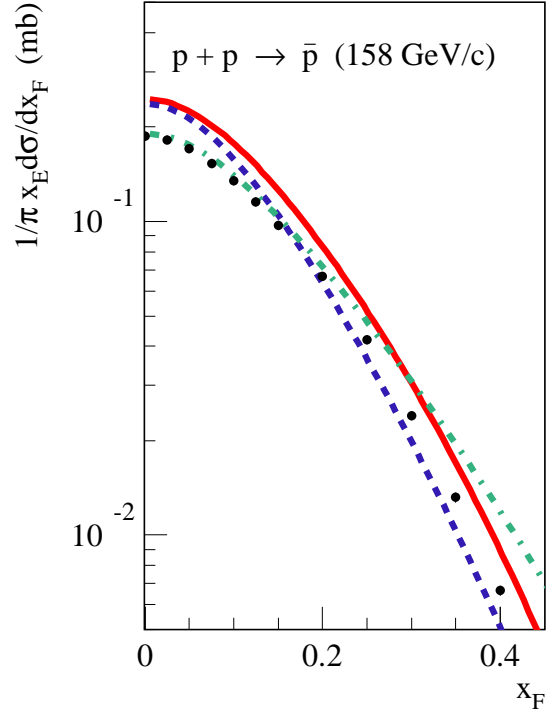}
}
\caption{Feynman $x$-spectra of $\bar p$'s (c.m.s.) in $pp$-collisions at $p_{\rm lab}=158$\,GeV/c, 
calculated using parameterizations by \citet{1983JPhG....9.1289T} and \citet{2003PhRvD..68i4017D}, 
in comparison with NA49 data \citep{2010EPJC...65....9A}; the lines are labeled as in Fig.~\ref{fig:ap19-ng}.\label{fig:ap158-ng}}
\end{figure}

\begin{figure}[p]
\center{
\includegraphics[width=0.45\textwidth]{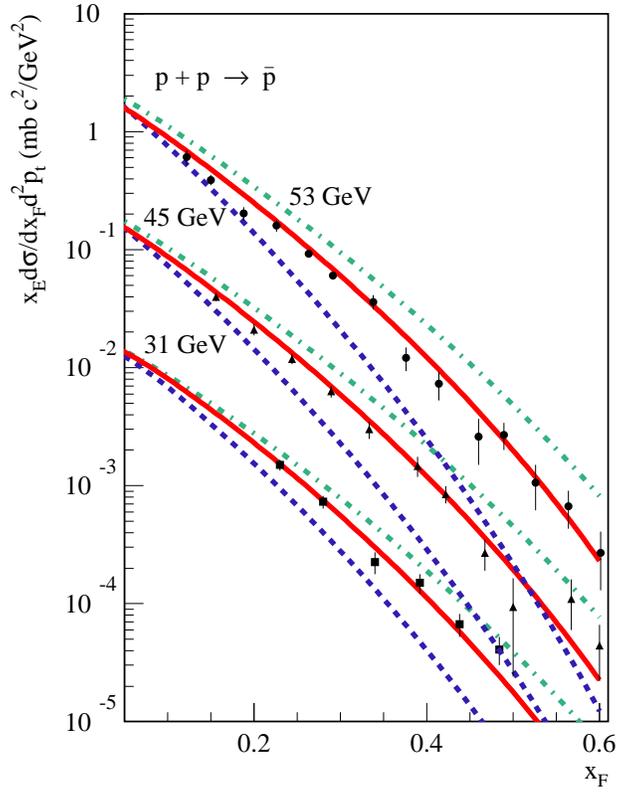}
}
\caption{Feynman $x$-spectra of $\bar p$'s for fixed c.m.s.\ angle $\tan \theta = 2.66/\sqrt{s}$ in $pp$-collisions
at $\sqrt{s}=31$ ($\times 10^{-2}$), 45 ($\times 10^{-1}$), and 53\,GeV:
parameterizations by \citet{1983JPhG....9.1289T} and \citet{2003PhRvD..68i4017D}
compared with experimental data \citep{1973NuPhB..56..333A};
the lines are labeled as in Fig.~\ref{fig:ap19-ng}.\label{fig:ap-sps-ng}}
\end{figure}

\begin{figure}[p]
\center{
\includegraphics[width=0.45\textwidth]{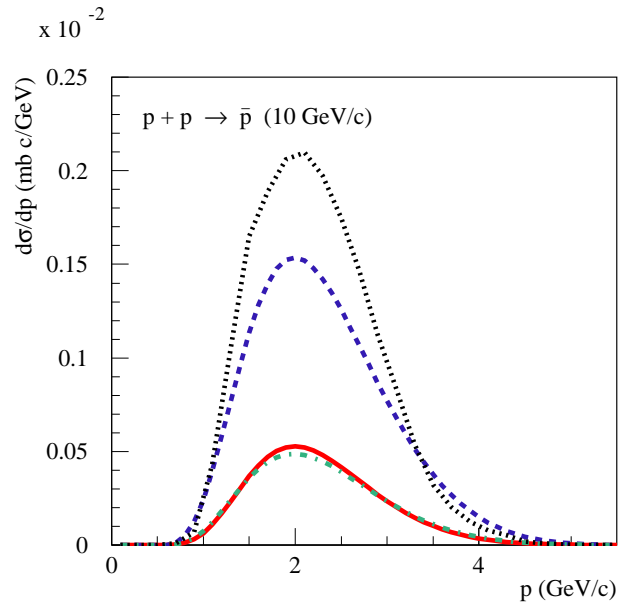}
}
\caption{The $p_{\rm t}$-integrated laboratory momentum spectra of $\bar p$'s 
in $pp$-collisions at $p_{\rm lab}=10$\,GeV/c calculated using parameterizations by \citet{1983JPhG....9.1289T} (solid, red)
and by \citet{2003PhRvD..68i4017D} (parameter set 1 -- dashed blue, parameter set 2 --
dot-dashed green) in comparison with QGSJET-IIm results (dotted, black).
\label{fig:ap10-ng}}
\end{figure}

To trace the source of the differences between QGSJET-IIm and the
parameterizations of \citet{1983JPhG....9.1289T} and \citet{2003PhRvD..68i4017D} in the predicted
$\bar p$-yields (Figs.~\ref{fig:zfac-pp}, \ref{fig:zdiff}),
 we compare the latter with selected accelerator data
in Figs.~\ref{fig:ap19-ng}, \ref{fig:ap158-ng}, and \ref{fig:ap-sps-ng}.
Starting with the results by \citet{1983JPhG....9.1289T}, we observe a 
generally reasonable agreement with the measured $\bar p$-spectra in $pp$-collisions over the wide energy range 
$E_p\sim 20-1500$\,GeV,\footnote{Comparing the parameterizations
of $\bar p $-production to the NA49 data in Fig.~\ref{fig:ap158-ng}, 
one has to take into account that the experimental spectrum has been
corrected for the contributions of $\bar \Lambda$ and $\bar \Sigma ^-$
decays,    which amounts to rescaling the results of 
 \citet{1983JPhG....9.1289T} and \citet{2003PhRvD..68i4017D}
 down by some 10 -- 15\%  \citep{2010EPJC...65....9A}.} which
dominates the $Z$-factors $Z^{pp}_{\bar p}$ for $E_{\bar p}^{\rm kin}$
between few GeV and few hundred GeV, i.e.\ where these factors
agree approximately with those calculated using QGSJET-IIm 
(see Figs.\ \ref{fig:zfac-pp}, \ref{fig:zdiff}). However, the 
 decrease of the $\bar p$-yield approaching the production
 threshold is faster in the parameterization of \citet{1983JPhG....9.1289T}
 than in QGSJET-IIm. This difference is illustrated in Fig.~\ref{fig:ap10-ng},
where the respective spectra for $pp$-collisions at 10\,GeV/c are compared.
 On the other hand, in the high energy asymptotics,
 the scaling-like behavior of the
 $\bar p$-spectra, implemented by \citet{1983JPhG....9.1289T},
 is broken both by the
 rise of $\sigma_{pp}^{\rm inel}$ and by the increase of the central
 rapidity density of secondary hadrons,
 as discussed in Section \ref{sec:zfactors}. 
 It is this scaling violation which leads to a steady energy rise
 of the predicted $Z$-factors $Z^{pp}_{\bar{p}}$ in
 QGSJET-IIm (c.f.\ Fig.~\ref{fig:zfac-pp}).
 In contrast, one obtains constant $Z^{pp}_{\bar{p}}$ values 
 using the parameterization of \citet{1983JPhG....9.1289T} at $E_p\gtrsim 100$\,GeV.

Let us now turn to the two parameterizations proposed by
 \citet{2003PhRvD..68i4017D}.
Apart from neglecting the energy rise of the inelastic cross section,
\citet{2003PhRvD..68i4017D} made questionable assumptions 
concerning the energy-dependence
of the spectral shape for $\bar p$-production.
In one case (parameter set 2), they assumed that the normalized
(per inelastic event) $\bar p$-production spectrum rises as a power of
energy: as $s^{\Delta}$, with\footnote{It is noteworthy that
this leads asymptotically to non-conservation of energy, because 
the energy fraction transferred to antiprotons would also rise as 
$s^{\Delta}$.}  $\Delta\simeq 0.25$.
 While a power-law
energy rise is indeed expected for hadron production in the central
rapidity region ($x_{\rm F}\simeq 0$) \citep{2011APh....35...98D}, the Feynman scaling
is known to hold approximately for the forward spectral shape 
\citep[see, e.g., a discussion in][]{2012PhRvD..86d3004K}. The outcome of this extreme
assumption overcompensates the neglected energy rise of
 $\sigma_{pp}^{\rm inel}$ and results in a too steep increase of
 inclusive $\bar p$-spectra, which is already visible in 
Fig.~\ref{fig:ap-sps-ng}, despite the narrow energy range covered by 
CERN ISR.
In turn, in the other case (parameter set 1), the authors assumed
a very strong scaling violation for the forward spectral shape, 
$\propto (1-x_{\rm R})^{{\rm const}\times \ln s}$ 
($x_{\rm R}\simeq 2E^*/\sqrt{s}$ at high energy,
 with $E^*$ being $\bar p$-energy in
c.m.s.). In combination with the constant $\sigma_{pp}^{\rm inel}$ assumed,
this leads to a drastic softening of the inclusive $\bar p$-spectra in
the forward direction, see Fig.~\ref{fig:ap-sps-ng}.

%%%%%%%%%%%%%%%%%%%%%%%%%%%%%%%%%%%%%%%%%%%%%%%%%%%%%%%%%%%%%%%%%%%%%%%
%%%%%%%%%%%%%%%%%% FIGURES & TABLES %%%%%%%%%%%%%%%%%%%%%%%%%%%%%%%%%%%%%%%%%%
%%%%%%%%%%%%%%%%%%%%%%%%%%%%%%%%%%%%%%%%%%%%%%%%%%%%%%%%%%%%%%%%%%%%%%%

\clearpage

%$E_{\bar p}^{\rm kin}$ & Primary & Target 
% &\multicolumn{6}{c}{$Z^{ij}_{\bar p}(E_{\bar p},\alpha)$, mb} \\
% (GeV)& nucleus& nucleus & $\alpha=2$ & $\alpha=2.2$ & 
% $\alpha=2.4$ & $\alpha=2.6$ &$\alpha=2.8$ & $\alpha=3$ 

\begin{deluxetable*}{rlcllllll}
\tablecolumns{9}
\tablewidth{0pc}
\tablecaption{$Z$-factors for ($\bar p+\bar n$) production,
$Z^{ij}_{\bar{p}}(E_{\bar{p}},\alpha)$, calculated with QGSJET-IIm
\label{tab:z-factors}}
\tablehead{
&
&
&
\multicolumn{6}{c}{$Z^{ij}_{\bar p}(E_{\bar p},\alpha)$, mb}\\
\colhead{$E_{\bar p}^{\rm kin}$, GeV} &
\colhead{Projectile nucleus} & 
\colhead{Target nucleus} & 
\colhead{$\alpha=2$} & 
\colhead{$\alpha=2.2$} & 
\colhead{$\alpha=2.4$} &
\colhead{$\alpha=2.6$} & 
\colhead{$\alpha=2.8$} &
\colhead{$\alpha=3$}
}
\startdata

1 & $p$  & $p$ & 
\phant 0.0254   & \phant 0.0138  & \phant 0.00772    & 0.00441   & 0.00258  &  0.00153\\

1 & He   & $p$ &
\phant 0.0808   & \phant 0.0442   & \phant 0.0248   & 0.0143   &  0.00838  & 0.00501\\

1 & CNO ($A=14$)   & $p$ &
\phant 0.184   & \phant 0.101   & \phant 0.0567   & 0.0326   &  0.0192  & 0.0115\\

1 & Mg-Si ($A=25$)   & $p$ &
\phant 0.273   & \phant 0.150   & \phant 0.0845   & 0.0486   &  0.0286  & 0.0171\\

1 & Fe ($A=56$)   & $p$ &
\phant 0.447   & \phant 0.245   & \phant 0.138   & 0.0792   &  0.0465  & 0.0278\\

1 &  $p$ & He   & 
\phant 0.0919   & \phant 0.0498   & \phant 0.0277   & 0.0158   &  0.00920    & 0.00546\\

1 &  He & He   & 
\phant 0.271  & \phant 0.147   & \phant 0.0824   & 0.0472   &  0.0276    & 0.0165\\

1 & CNO ($A=14$)   & He &
\phant 0.649  & \phant 0.352   & \phant 0.196   & 0.112   &  0.0654    & 0.0389\\

1 & Mg-Si ($A=25$)   & He &
\phant 0.933  & \phant 0.506   & \phant 0.282   & 0.161  &  0.0937    & 0.0556 \\

1 & Fe ($A=56$)   & He &
\phant 1.53   & \phant 0.834  & \phant 0.468   & 0.269   &  0.158  & 0.0944\smallskip\\

10 & $p$   & $p$ & 
\phant 0.279   & \phant 0.164  & \phant 0.100    & 0.0633   & 0.0413  &  0.0276\\

10 & He   & $p$ &
\phant  0.979   & \phant 0.573   & \phant 0.350   & 0.222   &  0.144  & 0.0964\\

10 & CNO ($A=14$)   & $p$ &
\phant 2.67   & \phant 1.58   & \phant 0.978   & 0.624   &  0.410  & 0.276\\

10 & Mg-Si ($A=25$)   & $p$ &
\phant 4.22   & \phant 2.50   & \phant 1.54   & 0.977   &  0.639  & 0.428\\

10 & Fe ($A=56$)   & $p$ &
\phant 7.78   & \phant 4.63   & \phant 2.87   & 1.84   &  1.21  & 0.815\\

10 &  $p$ & He   & 
\phant 0.970   & \phant 0.560   & \phant 0.339   & 0.213   &  0.138    & 0.0917\\

10 &  He & He   & 
\phant 3.16  & \phant 1.83   & \phant 1.11   & 0.695   &  0.449    & 0.298\\

10 & CNO ($A=14$)   & He &
\phant 9.16  & \phant 5.33   & \phant 3.24   & 2.04   &  1.32    & 0.875\\

10 & Mg-Si ($A=25$)   & He &
\phant 14.5  & \phant 8.45 & \phant 5.16   & 3.26  &  2.12    & 1.41 \\

10 & Fe ($A=56$)    & He &
\phant 26.0   & \phant 15.2   & \phant 9.31  & 5.90   &  3.85  & 2.57 \smallskip\\

100 & $p$   & $p$ & 
\phant 0.535   & \phant 0.308  & \phant 0.187    & 0.119   & 0.0789  &  0.0536\\

100 & He   & $p$ &
\phant 2.03   & \phant 1.17   & \phant 0.715   & 0.455   &  0.300 & 0.204\\

100 & CNO ($A=14$)   & $p$ &
\phant 6.21   & \phant 3.61   & \phant 2.20   & 1.40   &  0.926  & 0.628\\

100 & Mg-Si ($A=25$)   & $p$ &
\phant 10.6   & \phant 6.19   & \phant 3.79   & 2.42   &  1.60  & 1.09\\

100 & Fe ($A=56$)   & $p$ &
\phant 21.8   & \phant 12.7   & \phant 7.80   & 4.99  &  3.30 & 2.25\\

100 &  $p$ & He   & 
\phant 1.79   & \phant 1.02   & \phant 0.612   & 0.385   &  0.251    & 0.169\\

100 &  He & He   & 
\phant 6.41  & \phant 3.66   & \phant 2.21   & 1.39   &  0.914    & 0.619\\

100 & CNO ($A=14$)   & He &
\phant 20.8  &  \phant 11.9   &  \phant 7.15   & 4.50   &  2.94    & 1.98\\

100 & Mg-Si ($A=25$)   & He &
\phant 35.2  & \phant 20.2 &  \phant 12.2   & 7.69  &  5.04   & 3.40 \\

100 & Fe ($A=56$)    & He &
\phant 72.0  & \phant 41.3   & \phant 24.9  & 15.7   &  10.2  & 6.87 \smallskip\\
 
1000 & $p$   & $p$ & 
\phant 0.721   & \phant 0.410  & \phant 0.248    & 0.158   & 0.105  &  0.0715\\

1000 & He   & $p$ &
\phant 2.79   & \phant 1.60   & \phant 0.978   & 0.625   &  0.416 & 0.286\\

1000 & CNO ($A=14$)   & $p$ &
\phant 8.87   & \phant 5.12   & \phant 3.12   & 1.99   &  1.32  & 0.903\\

1000 & Mg-Si ($A=25$)   & $p$ &
\phant 15.3   & \phant 8.81   & \phant 5.37   & 3.42   &  2.27   & 1.55\\

1000 & Fe ($A=56$)   & $p$ &
\phant 31.4   & \phant 18.1   & \phant 11.1  & 7.09  &  4.72 & 3.24\\

1000 &  $p$ & He   & 
\phant 2.35   & \phant 1.32   & \phant 0.787   & 0.495   &  0.324    & 0.221\\

1000 &  He & He   & 
\phant 8.62  & \phant 4.85   & \phant 2.90   & 1.82   &  1.19    & 0.805\\

1000 & CNO ($A=14$)   & He &
\phant 29.0  &  \phant 16.4   & \phant 9.81   & 6.18   &  4.05    & 2.75 \\

1000 & Mg-Si ($A=25$)   & He &
\phant 48.1  &  \phant 27.0 &  \phant 16.0  & 10.0  &  6.50   & 4.36 \\

1000 & Fe ($A=56$)    & He &
\phant 103 & \phant 58.1   & \phant 34.9  & 22.1  &  14.5  & 9.85 \smallskip\\

10000 & $p$   & $p$ & 
\phant 0.931   & \phant 0.516  & \phant 0.307    & 0.193   & 0.127  &  0.0860\\

10000 & He   & $p$ &
\phant 3.60   & \phant 2.02   & \phant 1.20   & 0.754   &  0.493 & 0.334\\

10000 & CNO ($A=14$)   & $p$ &
\phant 11.5  & \phant 6.46   & \phant 3.87   & 2.44   &  1.61 & 1.09\\

10000 & Mg-Si ($A=25$)   & $p$ &
\phant 19.9   &  \phant 11.1   & \phant 6.61   & 4.12   &  2.68 & 1.80\\

10000 & Fe ($A=56$)   & $p$ &
\phant 41.6   & \phant 23.4   & \phant 13.9 & 8.72  &  5.69 & 3.84\\

10000 &  $p$ & He   & 
\phant 3.02   & \phant 1.65   & \phant 0.959   & 0.587   &  0.376    & 0.249\\

10000 &  He & He   & 
\phant 11.1  & \phant 6.14   & \phant 3.61   & 2.24  &  1.46    & 0.983\\

10000 & CNO ($A=14$)   & He &
\phant 37.4  &  \phant 20.5  &  \phant 12.0   & 7.40   &  4.77    & 3.20 \\

10000 & Mg-Si ($A=25$)   & He &
\phant 63.7  &  \phant 35.0 &  \phant 20.5  & 12.7  &  8.18   & 5.48 \\

10000 & Fe ($A=56$)    & He &
\phant 135 & \phant 74.5   & \phant 43.6  & 27.0  &  17.4  & 11.6

\enddata
\end{deluxetable*}

\begin{deluxetable}{cccccc}
\tablecolumns{6}
\tablewidth{0pc}
\tablecaption{Spectral parameterizations for groups of CR nuclei
\citep{2004PhRvD..70d3008H}
\label{tab:para}}
\tablehead{
& \multicolumn{5}{c}{Groups of nuclei}\\
\colhead{Parameters} & 
\colhead{H ($A$=1)} & 
\colhead{He ($A$=4)} & 
\colhead{CNO ($A$=14)} & 
\colhead{Mg-Si ($A$=25)} &
\colhead{Fe ($A$=56)}
}
\startdata
$K$ & 14900 & 600 & 33.2 & 34.2 & 4.45 \\

$\alpha$ & 2.74 & 2.64 & 2.60 & 2.79 & 2.68 
\enddata
\end{deluxetable}

\begin{deluxetable}{ccccc}
\tablecolumns{5}
\tablewidth{0pc}
\tablecaption{Energy dependence of the nuclear enhancement factor
$\epsilon_{\bar p}$ for CR composition
given in Table~\ref{tab:para}
\label{tab:enh-factor}}
\tablehead{
\colhead{$E_{\bar p}^{\rm kin}$, GeV } & 
\colhead{10} & 
\colhead{100} & 
\colhead{1000} & 
\colhead{10000}
}
\startdata
$\epsilon_{\bar p}$ & 1.71  &  1.85  &  2.00  &  2.16 
\enddata
\end{deluxetable}

\bibliography{references}
\end{document}